\documentclass[prl,twocolumn,superscriptaddress,
preprintnumbers,nofootinbib]{revtex4}
\usepackage{graphicx}
\usepackage{amsmath}
\usepackage{amssymb}
\usepackage{color}
\usepackage{hyperref}
\usepackage{url}
\usepackage{breakurl}
\usepackage{epstopdf} 
\newcommand{\be}{\begin{equation}}
\newcommand{\ee}{\end{equation}}
\newcommand{\bea}{\begin{eqnarray}}
\newcommand{\eea}{\end{eqnarray}}

\def\met{\slash{\!\!\!\!E}_{\text{T}}}
\def \red#1 {\textcolor{red}{#1}\ }

\begin{document}

\title{Measuring Di-Higgs Physics via the 
$t \bar t hh \to t \bar t  b\bar b b\bar b $ Channel}

\author{Tao Liu}
\email{taoliu@ust.hk}
\affiliation{Department of Physics, The Hong Kong University of  Science and Technology, \\Clear Water Bay, Kowloon, Hong Kong, P.R.China}

\author{Hao Zhang}
\email{zhanghao@physics.ucsb.edu}
\affiliation{Department of Physics, University of California, 
Santa Barbara, California 93106, USA}

\begin{abstract}
The measurement of di-Higgs physics can provide crucial 
information on electroweak phase transition 
in the early Universe and significant clues on new physics 
coupling with the Higgs field directly. 
This measurement has been suggested to be pursued mainly via 
the $pp \to hh$ production. In this letter,  we propose a new 
strategy to do that, i.e., via 
the $pp \to t \bar t hh $ production. Because of its positive correlation 
with the rescaled tri-Higgs coupling $\frac{\lambda}{\lambda_{\rm SM}}$ (in comparison to a negative one for the 
$pp \to hh$  production) in the neighborhood of $\frac{\lambda}{\lambda_{\rm SM}} \sim 1$, 
the $pp \to t\bar t hh$ production complements 
the $pp \to hh$ one in measuring di-Higgs physics, particularly for $\frac{\lambda}{\lambda_{\rm SM}}>1$, 
at both the High Luminosity LHC (HL-LHC) and a next-generation $pp$-collider.
As an illustration, we work on the process $pp\to t\bar thh \to t\bar t  b\bar b  b\bar b$. 
We show that  a statistical significance of $> 2.0 \sigma$ at the HL-LHC, comparable 
to that of the $pp \to hh \to b\bar b \gamma\gamma$ channel, and a statistical significance 
of $\sim 5 \sigma$ at a 100 TeV $pp$-collider, with 3000 fb$^{-1}$ of data, are achievable 
in searching for the di-Higgs production with $\frac{\lambda}{\lambda_{\rm SM}} = 1$.  
\end{abstract}


\maketitle

\section{Introduction}
\label{sec:intro}

The measurement of di-Higgs physics plays an important role 
in particle physics and cosmology. In the standard model (SM) of 
particle physics, the Higgs field induces electroweak (EW) 
phase transition (EWPT) as an order parameter, by interacting with itself, 
while the measurement of di-Higgs physics can provide information 
in this regard, given the involvement 
of the tri-Higgs coupling $\lambda$ in the di-Higgs production. 

This may have important implications in cosmology. As is well known, 
if the EWPT is of strong enough first order, the cosmic baryon asymmetry (CBA)
could be generated via EW baryogenesis. 
In the SM, however, such an EWPT requires 
a Higgs boson much lighter than 125 GeV, 
and hence can not be achieved without violating 
the current experimental bounds. 
To achieve this goal, several mechanisms were 
introduced in the SM extensions: 
(1) by introducing loop-corrections mediated by new scalar particles to the effective Higgs potential;  
(2) by incorporating nonrenormalizable operators such as $\frac{|H|^6}{\Lambda^2}$
in the effective Higgs potential; 
and (3) by mixing the Higgs field with a 
singlet scalar at tree level. No matter in which case, 
there is a strong correlation between the EWPT dynamics  
and the tri-Higgs coupling. As being pointed out in \cite{Noble:2007kk}, 
at zero temperature, the tri-Higgs coupling favored by the mechanisms (1) and (2) could be  
a couple of times larger than its SM value $\lambda_{\rm SM}$, while the   
tri-Higgs coupling favored by the mechanism (3) could be as small as 
$\sim 0.1 \lambda_{SM}$. 

In addition to the tri-Higgs coupling,  the 
di-Higgs production may receive contributions from new physics coupling with the Higgs field. 
The anomalous Higgs-top operator  $\frac{t\bar t hh}{\Lambda}$ is such 
an example, which often arises in the little Higgs or the composite Higgs 
models by integrating out a heavy top partner \cite{Dib:2005re,Grober:2010yv}. 
With an insertion into the gluon fusion loop, this operator can significantly modify the 
di-Higgs physics \cite{Dawson:2013uqa,Chen:2014xra,Chen:2014xwa}. 
The measurement of di-Higgs physics therefore provides a nice tool to probe both 
 the EWPT and new physics coupling with the Higgs field.

Because of this, the discovery of the Higgs boson in 2012 \cite{Aad:2012tfa,Chatrchyan:2012ufa} 
immediately motivated a series of studies on the 
measurement of di-Higgs physics at Large Hadron 
Collider (LHC) and High Luminosity LHC (HL-LHC) at which 
300 fb$^{-1}$ and 3000 fb$^{-1}$ of data are expected to be collected at each of 
the ATLAS and the CMS detecters, respectively~\cite{Dolan:2012rv,Shao:2013bz,
Barr:2013tda,deLima:2014dta,Barger:2013jfa,Papaefstathiou:2012qe,
Goertz:2013kp}, and at a next-generation $pp$-collider \cite{Yao:2013ika}. 
Currently the measurement of di-Higgs physics is 
suggested to be pursued mainly via:
\begin{itemize}

\item Channel 1~\cite{Dolan:2012rv,Baglio:2012np,Barr:2013tda,deLima:2014dta,Barger:2013jfa,Papaefstathiou:2012qe,Goertz:2013kp}: $pp \to hh \to  
b \bar b \gamma \gamma$, $b \bar b \tau \tau$, $b \bar b W W^*$

\end{itemize}
which provides the best sensitivity so far in measuring di-Higgs physics, 
though some preliminary work has also been done on~\cite{Dolan:2013rja}
\begin{itemize}

\item Channel 2: $pp \to jjhh$  

\end{itemize}
(for a recent review, see~\cite{Baglio:2014aka}). 
The production of both channels however has a negative dependence on
 $\frac{\lambda}{\lambda_{\rm SM}}$ in the SM neighborhood, 
 with the differential cross section $\frac{d (\sigma / \sigma_{\rm SM}) }{d (\lambda /  \lambda_{\rm SM})}$  
 becoming less and less negative as $\frac{\lambda}{\lambda_{\rm SM}}$ 
increases~\cite{Frederix:2014hta}. Both effects can suppress their sensitivities   
in measuring the tri-Higgs coupling, particularly if $\frac{\lambda}{\lambda_{\rm SM}}>1$. 
In addition, in both channels there exists a degeneracy of production cross section
with respect to $\frac{\lambda}{\lambda_{\rm SM}}$. Breaking this degeneracy may 
further suppress the sensitivities. To explore the di-Higgs physics, therefore, 
a complementary strategy  is needed, particularly in the parameter region with 
$\frac{\lambda}{\lambda_{\rm SM}}>1$. 

In this letter we propose a new strategy to explore the di-Higgs 
physics at $pp$-colliders, say, via  
\begin{itemize}

\item Channel 3: $pp \to t\bar t hh$.

\end{itemize}
The $t \bar t hh$ production has a cross section monotonically increasing with 
respect to $\frac{\lambda}{\lambda_{\rm SM}}$~\cite{Frederix:2014hta}, 
with the $\frac{d (\sigma / \sigma_{\rm SM}) }{d (\lambda /  \lambda_{\rm SM})}$  
 becoming more and more positive as $\frac{\lambda}{\lambda_{\rm SM}}$ 
increases~\cite{Frederix:2014hta}, which potentially enables it to 
fulfill our needs. A comparison of the cross sections between 
the $pp \to hh$ production and the $pp \to t\bar t hh$ production are provided in Table \ref{tab:xsec}. 
\begin{table}[!htb]
\caption{A comparison of the next-to-leading order (NLO) 
cross sections (in fb) of 
$t\bar t hh $  and $ hh$ at $pp$-colliders 
\cite{Frederix:2014hta}.}
\begin{center}
\begin{tabular}{ccc}
\hline\hline
$\sqrt{s}$  &
     $pp \to t\bar t hh $   &   $ pp \to hh $    \\ \hline
14 TeV & $0.981^{+2.3+2.3\%}_{-9.0-2.8\%}$ 
&  $34.8^{+15+2.0\%}_{-14-2.5\%}$ \\
100 TeV & $\sim90$ & $\sim 1200$      \\
\hline\hline
\end{tabular}
\end{center}
\label{tab:xsec}
\end{table}
Though its production cross section is an order smaller than the $pp \to hh$ one, 
the extra $t\bar t$ in the $t\bar t hh$ events may suppress one order or orders more 
backgrounds. So, the $t \bar t hh$ production opens a new avenue to measure 
di-Higgs physics at HL-LHC and a next-generation $pp$-collider, 
with the decays $ hh \to   b \bar b b \bar b$, $b \bar b \gamma \gamma$, $b \bar b \tau \tau$, $b \bar b W W^*$, $b \bar b ZZ^*$, etc.

As an illustration, we will focus on the $pp\to t\bar t hh$ production with $hh \to  b \bar b b \bar b$ at the HL-LHC, 
which results in a sensitivity comparable to that of the $pp \to hh \to b\bar b \gamma \gamma$ 
in searching for the SM di-Higgs production, and shortly discuss its sensitivity at a 100 TeV $pp$-collider.  
We would emphasize that this doesn't mean that, for the $t\bar t hh$ production, the $hh\to b\bar bb\bar b$    
has a better sensitivity at a 100 TeV $pp$-collider, compared to its other decay modes.

\section{Analysis Strategy}
\label{sec:pheno}
In the analysis of measuring di-Higgs physics 
via $pp\to t \bar t hh \to t \bar t b \bar bb\bar b$ at the HL-LHC, 
we allow the top pairs to decay either semi-leptonically or 
leptonically (with $\ell = e, \mu$). Unless indicated explicitly, 
the discussions below on our strategies can be applied to both cases. 
In the analyses, the main irreducible backgrounds  include
\begin{itemize}
\item $pp\to t\bar t b\bar bb\bar b $,
\item $pp\to t\bar t h b\bar b, h\to b\bar b $,
\item $pp\to t\bar t Z b\bar b, Z\to b\bar b $,
\end{itemize}
and the main reducible backgrounds include 
\begin{itemize}
\item $pp\to t\bar t b\bar bjj $,
\item $pp\to t\bar t h jj, h\to b\bar b $,
\end{itemize}
According to \cite{CMS:2013xfa},  
a $70\%$ $b$-tagging rate at 14 TeV LHC will lead to a $2\%$ mistag rate for light jets and 
a $24\%$ mistag rate for charm jets, with a 50 pile-up assumed. 
Thus only charm jets will be considered for reducible backgrounds. 
The contributions of $pp\to t(\bar t)+b{\text{-jets}}$, $W^\pm+b{\text{-jets}}$, $t\bar t h Z$, $t\bar t Z Z$ and $t\bar t Z jj$ 
to the backgrounds are negligibly smaller than that of the top-pair plus multi-jets events.
So they will not be considered. 

Our analysis framework is described in the following. We use MadGraph5 
 \cite{Alwall:2014hca} to generate leading-order (LO) signal and background 
events, with the CTEQ6L1 parton distribution function (PDF)
\cite{Pumplin:2002vw} applied. All of the signal and background 
events are showered by Pythia6.4 \cite{Sjostrand:2006za}.
We use DELPHES 3 \cite{deFavereau:2013fsa} for detector
simulations in which the $b$-tagging efficiency and the mistag 
rate of $c$-jets are
tuned to be $70\%$ and $24\%$, respectively.
To reconstruct the Higgs invariant mass, the   
energy of $b$-jets is rescaled by a factor 
\begin{eqnarray}
1.0+\frac{p_1}{p_T^{j_b}}+p_2
\end{eqnarray}
with $p_1=6.15298$, $p_2=-0.006007, $which is obtained
from the $Zb\to\mu^+\mu^-b$ process.

{\bf [Preselection]} 
In the analysis, electrons and muons are isolated by passing the cut 
\be
|\eta^\ell|<2.5,~I_{iso}<0.1.
\ee
Here $I_{iso}$ is energy accumulation (except the 
energy of the charged target lepton) in a  
$\Delta R=0.3$ cone around the charged target lepton 
which is rescaled by the lepton energy. In the semi-leptonic 
and di-leptonic top-pair cases, all events are required to   
contain exactly one isolated charged lepton 
with $p_T^\ell>20~{\text{GeV}}$, and exactly two isolated 
opposite-sign charged leptons with $p_T^\ell>10~{\text{GeV}}$,  respectively.

Jets are reconstructed by using anti-$k_T$ algorithm 
with $\Delta R=0.5$ within the $|\eta^j|<4.5$ region. They 
are considered for b-tagging if and only if they fall within the 
tracker acceptance of $|\eta^j| < 2.5$. We require at least
7 jets with $p_T^j>20$ GeV in the semi-leptonic top-pair case 
and at least 5 jets with $p_T^j>20$ GeV in the di-leptonic top-pair case. 
In addition, we require at least 5 of them be tagged as $b$-jets. 
A cut for missing transverse energy $\met >$ 30 GeV 
is applied in the semi-leptonic top-pair
case and the leptonic top-pair case with a pair of charged leptons of the same flavor. 
In the latter case, we require the di-lepton invariant mass satisfy
\be
|m_{\ell\ell}-m_{Z}|>10~{\text{GeV}},
\ee
with $m_Z=91.1876$ GeV, to suppress the background 
events which contain a pair of charged 
leptons due to $Z$-boson decay. 
In the leptonic top-pair case with a pair of charged leptons of different flavors
(i.e., $e\mu$), no $\met$ cut and $Z$ mass window 
cut will be applied. 

{\bf [Reconstruction of di-Higgs resonances]} 
The  two Higgs resonances are reconstructed 
by using $b$-jets in each event. 
Generically there is a combinatorial problem, due to the fact that   
top quarks decay into a bottom quark and a $W$ boson. 
To reconstruct the two Higgs resonances, 
we choose a combination $((b_1,b_2),(b_3,b_4))$ among 
the tagged $b$-jets which gives the minimum of
\be
\chi_h\equiv\sqrt{\left(\frac{m_{b_1b_2}-m_h}{\sigma_h}\right)^2
+\left(\frac{m_{b_3b_4}-m_h}{\sigma_h}\right)^2}.
\ee
Here $m_h=125.4$ GeV and $\sigma_h=30$ GeV are assumed.

\begin{figure}[!htb]
\includegraphics[width=.9\columnwidth]{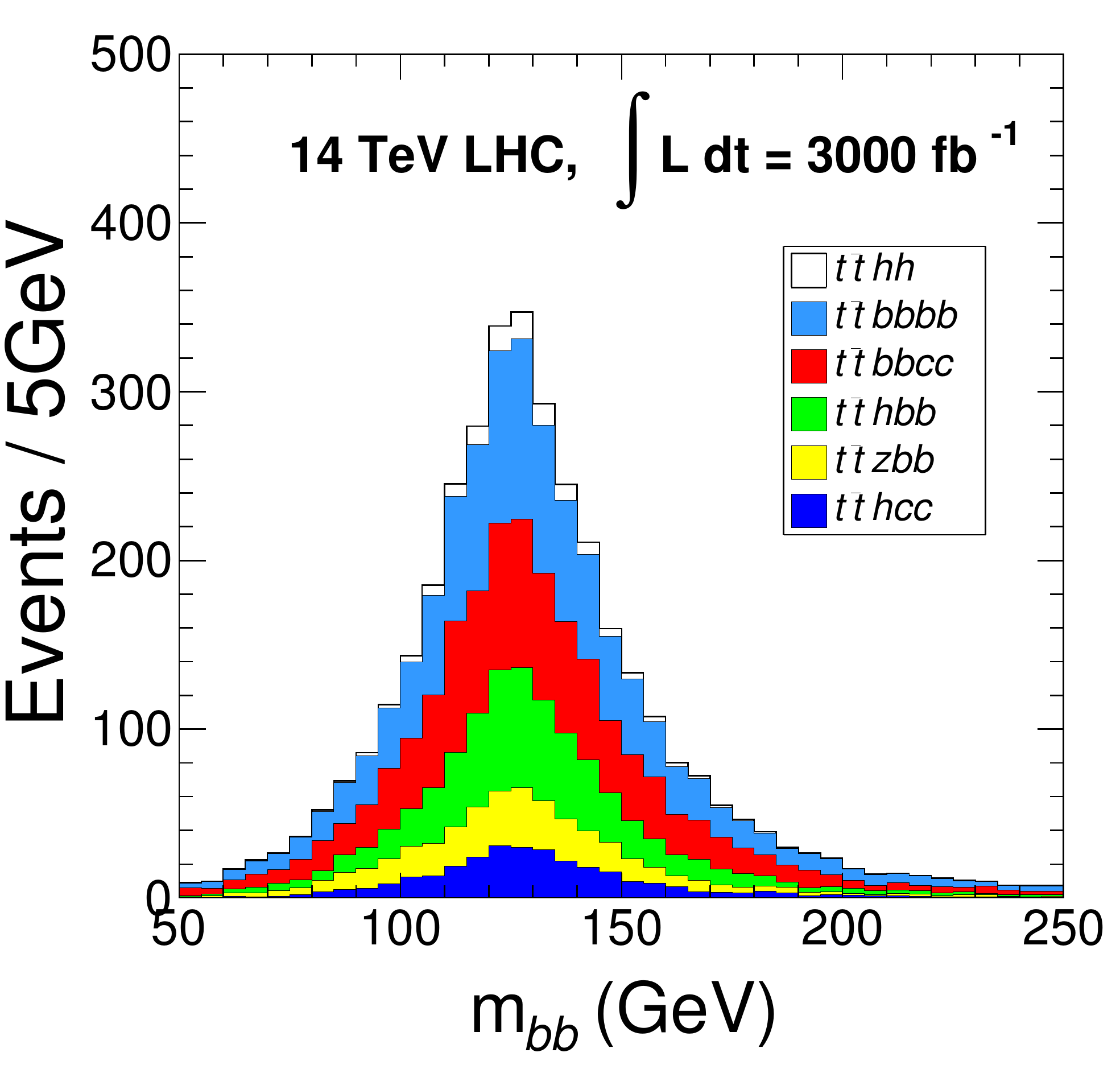}
\caption{The $m_{bb}$ reconstruction for the two $b\bar b$ pairs  which 
minimize the $\chi_h$ in each event after the preselection cut in the semileptonic 
top-pair case. 
\label{fig:mbb} }
\end{figure}

All events are required to pass the reconstruction cut of 
the di-Higgs resonances
\be
\chi_h<1.8.
\ee
In addition, one of the two selected $b$-jet pairs may have an 
invariant mass close to $m_h$ accidentally. 
In such a case, this cut may lose its effect since the second 
$b$-jet pair is allowed to have a relatively large deviation from 
$m_h$. To increase the cut efficiency, we require 
\be
\left(\frac{m_{b_1b_2}-m_h}{\sigma_h}\right)^2~
{\text{and}}~ \left(\frac{m_{b_3b_4}-m_h}{\sigma_h}\right)^2
\ee
be symmetric and neither of them is allowed to be larger than 1.9. 
As a result, all of the signal and background events surviving 
of this cut should have two $b$-jet pairs, with their invariant  
masses deviating from $m_h$ in a comparable way. 

{\bf [Reconstruction of top quark resonance]} 
To suppress the backgrounds with no top quarks,
we may reconstruct one of the top quarks in the signal events. 
In semi-leptonic top-pair case, we reconstruct the leptonic top quark 
by using the charged lepton ($\ell$), 
missing transverse energy ($\nu$) and a reconstructed jet ($j$). 
Here the jet is not necessary to be $b$-tagged, 
but it should not be any one among $b_1, b_2, b_3$ and $b_4$. 
The neutrino momentum 
along the beam-line direction is solved by using the 
$W$-boson mass-shell equation. Due to smearing effects, 
an imaginary solution is possible. So we require
 at least one real solution. For the events which have two real solutions, 
 we use both of them to calculate $m_{j\ell\nu}$. Then all events are required to  
pass a top-mass cut
\be
{\text{min}}|m_{j\ell\nu}-m_t|<50 \text{GeV} ,
\ee
where $m_t=173.2$ GeV. In spite of this, the cut for 
top-quark reconstruction might be too aggressive, given that  
the dominant non-top background $W + $jets is negligibly small 
due to the suppression by the requirement of at least seven 
 jets with at least five of them $b$-tagged in each event. 
 So in the next section, we will present the analysis 
results, both with and without top-quark reconstruction. 

As for the di-leptonic top-pair case, the main non-top backgrounds are
Drell-Yan process for the $ee$ and $\mu\mu$ channels, and di-boson+jets 
for the $e\mu$ channel, which are also sub-dominant. 
So, no top reconstruction will be applied in the di-leptonic top-pair case.

\section{Simulation Results}
\label{sec:results}

The cut flows of the signal and the background events 
in the semi-leptonic top-pair and the di-leptonic top-pair cases 
 are summarized in Table \ref{tab:14cutflow} and Table \ref{tab:14cutflow_dil}, respectively. 
 We find that the background contributed by faked $c$-jets is 
 important and hence is not negligible. The cut flows indicate that 
 a statistical significances as large as $S/\sqrt{B}=2.0\sigma$ (no top-quark reconstruction)
and $S/\sqrt{B}=1.5\sigma$ (with top-quark reconstruction) can be 
achieved in the semi-leptonic top-pair case. As for the di-leptonic top-pair case, 
the statistical significance is $S/\sqrt{B}=0.8\sigma$ at the HL-LHC. The sensitivities in 
these two analyses can be combined quadratically, which gives 
a sensitivity of $2.2\sigma$ (no top-quark reconstruction), in comparison to   
a statistical significance of $2.3\sigma$ expected  
to be achieved at the HL-LHC via $pp \to hh \to b\bar b \gamma\gamma$,   
with a $75\%$ $b$-tagging efficiency assumed~\cite{Yao:2013ika}. 

\begin{table}[h]
\caption{Cut flows of searching for 
$t \bar t hh \to t \bar t b \bar bb\bar b$ at the HL-LHC via the  
semi-leptonic top-pair channel. 
The unit used in the table is attobarn.}
\begin{center}
\begin{tabular}{ccccccc}
\hline\hline
$\sqrt{s}=14$ TeV 
& $t \bar t h h$ & $t \bar t b \bar bb\bar b$     
& $t \bar t b \bar bc\bar c$ 
&  $t \bar t h b\bar b$  & $t \bar t Z b\bar b$  & $t \bar t hc\bar c$  
 \\ \hline
Preselection &  39.0 & 390.6 & 353.1 & 222.7 & 126.8 &98.2  \\ 
Di-Higgs rec. & 33.0 & 269.3 & 242.1 & 171.0 & 93.5 &76.8  \\ 
Top rec. & 19.5 & 160.7 & 149.0 & 102.8 & 54.6 &47.1   \\ \hline\hline
\end{tabular}
\end{center}
\label{tab:14cutflow}
\end{table}

\begin{table}[!htb]
\caption{Cut flows of searching for 
$t \bar t hh \to t \bar t b \bar bb\bar b$ at the HL-LHC via the 
dileptonic top-pair channel. The unit used in the table is attobarn.}
\begin{center}
\begin{tabular}{ccccccc}
\hline\hline
$\sqrt{s}=14$ TeV 
& $t \bar t h h$ & $t \bar t b \bar bb\bar b$     
& $t \bar t b \bar bc\bar c$ 
&  $t \bar t h b\bar b$  & $t \bar t Zb\bar b$  
&  $t \bar t hc \bar c$ \\ \hline
Preselection  &  4.8 & 41.6 & 30.6 & 22.6 & 9.7 & 8.1  \\ 
Di-Higgs rec. & 4.1 & 27.1 & 20.7 & 16.8 & 7.4 & 6.4  \\ \hline\hline
\end{tabular}
\end{center}
\label{tab:14cutflow_dil}
\end{table}

\section{Discussions}
\label{sec:measure}

Leading-order discussions have been pursued, 
regarding the measurement of di-Higgs physics via the 
$t \bar t hh$ channel. In the illustrational case with $pp \to t \bar t hh\to t \bar t b\bar b b\bar b$, 
we show that a sensitivity  comparable to that of the $pp \to hh \to b \bar b \gamma\gamma$ 
channel is achievable in searching for the SM di-Higgs production 
at the HL-LHC, which is very encouraging. However, we need to note that 
the dominant backgrounds in this case are $t\bar tb\bar bb\bar b$
and $t\bar tb\bar b jj$, both of which have a cross section of order $\alpha
^6_S$ at tree level. This may lead to a large theoretical uncertainty 
in estimating the backgrounds. A calculation of higher-order corrections 
therefore is important for suppressing this uncertainty. 
Alternatively, a data-driven method may help in this regard. 

At analysis level, a further improvement is certainly possible. For example, 
we may introduce color-flow variables such as the ``pull angle'' of $b$-jet 
pairs~\cite{Gallicchio:2010sw} to reconstruct the di-Higgs resonances, which has been 
shown to be useful in suppressing combinatorial backgrounds 
of multiple b-jets in both supersymmetric~\cite{Berenstein:2012fc} 
and non-supersymmetric~\cite{Li:2013xba} contexts.  In addition, 
we can apply more advanced analysis tools, such as the tool of jet-substructure 
and the multivariate method of Boost Decision Tree, which have been successfully 
applied for measuring di-Higgs physics in the channels 
$pp \to hh \to b \bar b \tau \tau$~\cite{Dolan:2012rv} 
and $b \bar b WW$~\cite{Papaefstathiou:2012qe}, respectively.

More importantly, the $pp \to t \bar thh$ provides a series of new opportunities 
to study di-Higgs physics at a next-generation $pp$-collider, 
with the decays $ hh \to   b \bar b b \bar b$, $b \bar b \gamma \gamma$, 
$b \bar b \tau \tau$, $b \bar b W W^*$, $b \bar b ZZ^*$, etc.
Though its production cross section is an order smaller than that 
of $pp \to hh$, the extra $t\bar t$ in the $t\bar t hh$ events 
may suppress one order or orders more backgrounds. As an illustration, let's consider the specific process 
$pp\to t\bar t hh \to   t\bar t  b \bar b b \bar b$ again at a 100 TeV $pp$-collider, 
with $t\bar t$ decaying semi-leptonically. Note, this doesn't mean that it has a better sensitivity 
compared to the other $hh$ decay modes in the $pp \to t\bar thh$ production. 
In this case, we modify the $p_T$ cuts for jets to be $p_T^j > 40$ GeV and 
the $\met$ cut to be $\met>50$ GeV, and   
require at least one jet with its $p_T$ greater 
than 100 GeV and at least one $b$-jet with its $p_T$ greater 
than 120 GeV. To reconstruct the di-Higgs resonances, we redefine $\chi_h$ to be 
\be
\chi_h \equiv \left[\left(\frac{m_{b_1b_2}-m_h}{\sigma_h}\right)^p
+\left(\frac{m_{b_3b_4}-m_h}{\sigma_h}\right)^p\right]^{1/p} .
\ee
We require the combination of $b$-jet pairs with the minimal $\chi_h$
satisfy $\chi_h < 2.5$ for $p=1.5$ and $\chi_h > 1.5$ for $p=0.2$. 
The latter is applied to avoid accidental ``di-Higgs'' resonances 
in the backgrounds. In addition, we require the di-Higgs invariant mass 
$m_{hh}< 750$ GeV, and $\left(|\Delta R_{b_1b_2}|^p-|\Delta R_{b_3b_4}|^p\right)^{1/p} < 0.1$
for $p=0.3$. The cut flows of both the signal and backgrounds are 
present in Table~\ref{tab:100cutflow}, which indicate  
a statistical significances $S/\sqrt{B}=4.9\sigma$ 
(no top-quark reconstruction)
and $3.3\sigma$ (with top-quark reconstruction) 
for 3ab$^{-1}$ of data.

\begin{table}[!htb]
\caption{Cut flows of searching for 
$pp \to t \bar t hh \to t \bar t b \bar bb\bar b$ at the 100 TeV $pp$
-collider via the  
semi-leptonic top-pair channel.  The unit used in the table is attobarn.}
\begin{center}
\begin{tabular}{ccccccc}
\hline\hline
$\sqrt{s}=100$ TeV 
& $t \bar t h h$ & $t \bar t b \bar bb\bar b$     
& $t \bar t b \bar bc\bar c$ 
&  $t \bar t h b\bar b$  & $t \bar t Zb\bar b$  
&  $t \bar t hc \bar c$
\\ \hline
Preselection       &  830.5 & 72678.7 & 13322.6 & 10231.8 & 3252.0 & 1995.7  \\ 
Di-Higgs rec.   & 608.4 & 31679.7 & 6285.2 & 5689.9 & 1504.0 & 1193.3  \\ 
Top rec. & 240.1 & 10384.4 & 2189.1 & 2208.6 & 428.0 & 384.9   \\ \hline\hline
\end{tabular}
\end{center}
\label{tab:100cutflow}
\end{table}

One application of the di-Higgs measurement is to probe the tri-Higgs coupling. 
A rough estimation based on the calculation in~\cite{Frederix:2014hta} 
gives 
\begin{eqnarray}
\frac{d (\sigma / \sigma_{\rm SM}) }{d (\lambda /  \lambda_{\rm SM})} \Big|_{\frac{\lambda}{\lambda_{\rm SM}}=1}  \sim 0.3 \label{dx14}
\end{eqnarray}
for the $pp \to t\bar t hh$ production, in comparison to its value $\sim -0.8$ for the $pp \to hh$ production~\cite{Yao:2013ika},
at a 14 TeV $pp$-collider. According to the analyses above, the SM tri-Higgs coupling 
can be measured with a statistical accuracy of  $\sim 150\%$ at the HL-LHC, 
and of $\sim 70\%$ at a 100 TeV $pp$-collider with 3ab$^{-1}$ of data 
(with the relation in Eq. (\ref{dx14}) assumed), via the channel 
$pp \to t\bar t hh \to t \bar t b \bar bb \bar b$. Here the former is 
based on a combination of the semi-leptonic and leptonic $t \bar t$ decay modes, 
and the latter is based on the semi-leptonic one only.  
Though the accuracy of this measurement 
is lower than what can be achieved at the HL-LHC via 
the $pp \to hh \to b \bar b \gamma\gamma$ channel, 
say, $\sim 50\%$~\cite{Yao:2013ika}, we are able to use it 
to preliminarily probe the $\frac{\lambda}{\lambda_{\rm SM}}$ shift required 
for generating strong enough first-order EWPT in the early Universe~\cite{Noble:2007kk}. 

The story could be more subtle if $\frac{\lambda}{\lambda_{\rm SM}} > 1$.  
Different from the $pp \to hh$ (similar for the $pp \to jjhh$) production whose cross section negatively  
depends on $\frac{\lambda}{\lambda_{\rm SM}}$ 
in the SM neighborhood, the $pp \to t \bar t hh$ production has a cross section
monotonically increasing with respect to $\frac{\lambda}{\lambda_{\rm SM}}$. 
Any positive shift in $\frac{\lambda}{\lambda_{\rm SM}}$ caused by new physics, such as 
the operator $\frac{|H|^6}{\Lambda^2}$ used for strengthening the EWPT in the early Universe~\cite{Noble:2007kk},
will lead to a suppression of the $pp \to hh$ production, and simultaneously 
an enhancement of the $pp \to t\bar t hh$ one, in this neighborhood.  
Meanwhile, the $\left |\frac{d (\sigma / \sigma_{\rm SM}) }{d (\lambda /  \lambda_{\rm SM})}\right |$  
becomes smaller for the $pp\to hh$ production and larger for 
the $pp\to t \bar t hh$ production as $\frac{\lambda}{\lambda_{\rm SM}}$ 
increases, which also leads to a suppression for the $pp\to hh$ sensitivity,   
and a simultaneous enhancement of the $pp\to t \bar thh$ sensitivity,   
in measuring the tri-Higgs coupling. For example~\cite{Frederix:2014hta}, 
with a shift 0.5 in $\frac{\lambda}{\lambda_{\rm SM}}$, 
the $pp \to t\bar thh$ production is enhanced by twice, relative to the $pp\to hh$ one, 
while the $\left |\frac{d (\sigma / \sigma_{\rm SM}) }{d (\lambda /  \lambda_{\rm SM})}\right |$
becomes comparable for both. This leads to a $pp \to t\bar thh \to t \bar t b\bar b b\bar b$ sensitivity roughly twice 
better than the  $pp \to hh \to b \bar b \gamma\gamma$ one in measuring the tri-Higgs coupling at the HL-LHC. 
Even worse, there exists a degeneracy of cross section
with respect to $\frac{\lambda}{\lambda_{\rm SM}}$ for the $pp \to hh$ production~\cite{Frederix:2014hta}. 
Breaking this degeneracy may further suppress its sensitivity. 
Given these considerations, the $pp \to t\bar t hh$ production may play a crucial role in measuring the tri-Higgs coupling 
and hence in exploring the CBA puzzle.  

Another application of the di-Higgs measurement 
is to search for new physics coupling with the Higgs field directly.  
The $t \bar t hh$ (including $t \bar t hh + \met$) production 
extensively exists in the scenarios of new physics. For example,   
it can be initiated by the pair production 
of top partners, in both supersymmetric (e.g., see~\cite{Berenstein:2012fc}) 
and non-supersymmetric (e.g., see~\cite{Li:2013xba}) contexts. 
In addition, higher dimensional operators in low-energy effective theories
may modify the $pp \to t \bar t hh$ production. $\frac{t \bar t hh}{\Lambda}$ is 
such an example which can contribute via top-quark pair production 
\cite{Dawson:2013uqa,Chen:2014xra,
Chen:2014xwa}. However, to achieve the double goals of  
measuring the tri-Higgs coupling and searching for 
new physics coupling with the Higgs field simultaneously, 
the $pp \to t\bar thh$ events need to be disentangled.

Based on the leading-order discussions above, we conclude that the $pp \to t\bar t hh$ channel 
opens a new avenue to measure di-Higgs physics, complementary 
to the channels $pp \to hh, jjhh$ suggested in the past, at both the 
HL-LHC and a next-generation $pp$-collider. A systematical exploration 
along this line is definitely required, which we will leave to a future work.  \\

\begin{acknowledgments}
{\bf [Acknowledgments]} T. L. is supported by his start-up fund at the 
Hong Kong University of Science and Technology. 
H. Z. is supported by the U.S.  DOE under 
Contracts No. DE-FG02-91ER40618 and DE-SC0011702. 
T. L. would like to thank Jing Shu and Lian-tao Wang for useful discussions, 
and to acknowledge the hospitality of the Aspen 
Center for Physics (Simons Foundation), 
where part of this work was completed.
H. Z. would like to thank Hua Xing Zhu for 
discussions. 
\end{acknowledgments}  

\bibliographystyle{apsrev}
\bibliography{draft_v7}
  
\end{document}